\RequirePackage{fix-cm}
\documentclass[smallextended]{svjour3}       % onecolumn (second format)
\smartqed  % flush right qed marks, e.g. at end of proof
\usepackage{graphicx}
\usepackage{amsmath,amsfonts,amssymb}
\usepackage{dsfont}
\usepackage{mathrsfs} %!PN
\allowdisplaybreaks[1]
\usepackage[a4paper]{geometry}
\begin{document}

\title{Weight Try-Once-Discard Protocol-Based $L_{2}$-$L_{\infty}$ State Estimation for Markovian Jumping Neural Networks with Partially Known Transition Probabilities%\thanks{Grants or other notes
}


\author{Cong Zou  \and Wei Chen \and   \and  %etc.
}


\date{Received: date / Accepted: date}

\maketitle

\begin{abstract}
%This paper discusses the $H_\infty$ state estimation issue in regard to Markovian jumping neural networks (MJNNs) under the scheduling of the Round-Robin protocol (RRP). The model takes into account mixed time-delays, sensor nonlinearities and exogenous disturbances, making it relatively general and comprehensive. The transmission of MJNNs signals invoked a communication scheme in which the RRP is used for the data transmissions in order to avoid undesirable data collisions. Protocol-dependent state estimator modeling of a hybrid switching system with mixed time delays and disturbances is designed for the first time to achieve asymptotic tracing for the neuron state. Using the Lyapunov stability theory and several asymptotic methods, sufficient conditions for guaranteeing the asymptotic stability of the state estimation are established under the constraint  of $H_\infty$ performance. By employing a combination of matrix analysis techniques, the estimator gain matrices are calculated by the feasible solutions to the linear matrix inequalities (LMIs). Finally, a numerical example and related simulations demonstrate the validity of the proposed model.

\end{abstract}

\section{Introduction}\label{sec:1}
For nearly decades, it is generally known that artificial neural networks (ANNs) have been continuously promoted and improved in different fields including signal transmission, memory storage and image recognition.
However, in practical engineering applications, actuator failure, parameter drift and changes in internal connections of subsystems often occur in actual control network systems, and these negative situations will randomly affect the structure of the control system. Therefore, the Markovian jumping neural networks (MJNNs) have been put forward and widely applied by more and more experts because of its effective modeling capability, which is a random switching system with multiple modes.  Considerable related research conclusions have been shown in  \cite{Li-Zuo-Wang,Daafouz-Riedinger-Iung,Hua-Zheng-Deng,Liang-Dai-Shen-Wang} and the references therein.

As is well known, transition probability determines the operation of the whole MJNNs. In an ideal situation, the analysis and design of the system are simple and convenient. Therefore, a large number of existing studies on MJNNs are obtained when the transition probability is fully known.
In practical, when the system is modeled as MJNNs, it is difficult to fully know the transition probability of the system jumping between modes according to great difficulty and high cost.
Therefore, instead of spending a lot of manpower and material resources to obtain the complete transition probability, from the control point of view, it is better to further study the more general MJNNs whose transition probability is partially unknown. In recent years, many studies based on it have been pay attention, see \cite{Yao-Liu-Lu-Xu-Zhou,Shen-13,Cui-Sun-Fang-Xu-Zhao,Yin-Shi-Liu-Liu,Zhang-B-L,Wang-1}.

In engineering applications, it is very critical to get the real state of the system for practical applications such as image recognition. It deserves to note that, the ability to obtain total state information is not enough due to physical limitations and technical difficulties. Consequently, state estimation method has been more and more recognized and studied in engineering control, which gives important rein to obtain real neuron state.
As for state estimation problem for MJNNs has recently become a hot topic of research with a great number of results reported in the literature, see \cite{Ji-Zhang-T,Liu-Shen-Li,Wang-Wang-Liu10,Zhao-Wang-Zou-Liu,Liu-Wang-Fei-Dong} and the reference therein. For instance,

On the other side, in the zone of state estimation,
it is often the case that the system performance should be considered to deal with the exogenous disturbances.
For unknown but energy bounded disturbances, one can resort to the $H_\infty$ and  $L_{2}$-$L_{\infty}$ performance indexes.
Regarding that the $L_{2}$-$L_{\infty}$ state estimation is able to constrain the effect from the energy-bounded external disturbances with unknown statistical characteristic on the estimation error.
On ground of the above analysis, plenty of research results have been obtained in the analysis of $L_{2}$-$L_{\infty}$ state estimation issue for MJNNs, see \cite{Chen-Zheng,Shen-Wang-Shen,Qian-Chen-Liu-Fe,Qian-Li-Chen-Yang,Qian-Li-Zhao-Chen,Shen-Xing-Wu-Cao-Huang}. However, in this paper the estimation problem for partially unknown transition probabilities in MJNNs are rarely mentioned and studied.

On another research hotspot is communication protocols, which can orchestrate the transmission order of sensor nodes. In major existing literatures about the estimation issue of MJNNs, there is an assumption that sensor nodes can receive transmission signals from the network at the same time. Nonetheless, it is usually impractical for networked systems because practical networks inevitably are subject to limited bandwidth which lead to data conflicts when encountering multiple information accesses.
In contrast with those traditional approaches without protocol scheduling, the introduction of communication protocol would cause some fundamental challenges to dynamic analysis. Up to now, there have been some elementary results about the MJNNs constrained by communication protocols including Round-Robin protocol (RRP), stochastic communication protocol (SCP) and weight try-once-discard protocol (WTOD protocol), see \cite{Li-Wang-Dong-Fei,Zou-Wang-Han-Zhou,Long-P-Ye,Liu-Wang-Ma-Zhang-Bo,Zou-Wang-Gao,Zhang-Peng,Shen-Wang-Shen-F,Ju-Wei-Ding-Liu,Aslam-Dai,Chen-Hu-Yu-Chen-Du}.
As such, the WTOD protocol is a kind of quadratic protocols, which differs from the periodic allocation of RRP. It can dispatch the transmission instants to certain sensor nodes based on a given quadratic selection principle. For instance, in \cite{Aslam-Dai}.
Therefore, it is more challenging to develop the communication protocol approach to address the Chen-Hu-Yu-Chen-Du $L_{2}$-$L_{\infty}$ state estimation problem for delayed MJNNs with partially known transition probabilities.

In view of the above-mentioned discussion, our intention is on the  $L_{2}$-$L_{\infty}$ state estimation issue for a class of delayed MJNNs with partially known transition probabilities based on the WTOD protocol. The elementary contributions of this paper are outlined as follows:
\begin{description}
  \item[1]It was the $L_{2}$-$L_{\infty}$ performance index that for the first time is initiated into the discussion on state estimation of delayed MJNNs with with partially known transition probabilities, which provides a more general promotion for the estimation error.
  \item[2] The WTOD protocol is adopted to dispatch the sensor nodes so as to effectively alleviate the updating frequency of output signals.
  \item[3] The hybrid effects of the time delays, Markov chain, and protocol parameters are apparently reflected in the co-designed estimator which can be solved by a combination of comprehensive matrix inequalities.
\end{description}

$\mathbf{Notations}.$ The notation used in this paper is normative.
There are the $n$-dimensional Euclidean space $\mathbb{R}^n$ and the sets of positive integers $\mathbb{Z}^+$;
$\|\cdot\|$  means the Euclidean norm; $I$, $0$ represent the identity matrix,  and zero matrix with appropriate dimensions, respectively.
Let $\mathbb{E}\{ \cdot \}$ is expressed as the mathematical expectation.
$P>0$ stands for that $P$ is a real symmetric positive definite matrix.
$\text{diag}\{\cdots\}$ denotes a block diagonal matrix.
 We use the notation $*$ as the ellipsis of symmetry terms.
 $\otimes$ represents the Kronecker product.

\section{Model Prescription and Lemmas}\label{sec:2}
\subsection{MJNNs Establishment }\label{sec:2.1}
We consider the discrete-time MJNNs with time-varying delays and energy-bounded exogenous disturbances described as follows:
\begin{eqnarray}\label{c01}
\left\{
\begin{array}{cll}
x(k+1)&=A(r_k)x(k)+B(r_k)f(x(k))+C(r_k)f(x(k-\tau(k))+D_1(r_k)\omega(k)\\
y(k)&=E(r_k)x(k)+D_2(r_k)v(k)\\
z(k)&=M(r_k)x(k)\\
\end{array}
\right.
\end{eqnarray}
where $x(k)=[x_{1}(k)\ \ \cdots\ \ x_n(k)]^{T}\in \mathbb{R}^{n}$ is the neuron state vector, $y(k)=[y_1(k)\ \ \cdots \ \ y_m(k)]^{T}\in \mathbb{R}^m$ is the measurement output of NNs, and $z(k)\in \mathbb{R}^q$ is the output vector to be estimated. $f(x(k)): \mathbb{R}^n\longrightarrow \mathbb{R}^n$ is nonlinear activation functions.  $A(r_k)$, $B(r_k)$,  $C(r_k)$, $D_1(r_k)$, $D_2(r_k)$, $E(r_k)$ and $M(r_k)$ are matrices with appropriate dimensions, mode-dependent time delay $\tau(k)$ satisfies $0<\bar{\tau}\leq\tau(k)\leq\tau$, and the exogenous disturbance input $\omega(k), v(k)\in \mathbb{R}^r$ satisfy $\omega(k),v(k)\in l_2[0,\infty]$.\\
The stochastic process $\{r_k,k\geq0\}$ is described by a discrete-time homogeneous Markov chain, which takes values in a finite state space $S=\{1,2,\cdots,N\}$ with the following mode transition probabilities:
\begin{displaymath}\label{c02}
\text{Prob}\{r_{k+1}=j|r_k=i\}=\pi_{ij}
\end{displaymath}
where $\pi_{ij}\geq0$, $\forall$ $i,j\in S$, and $\sum_{j=1}^N\pi_{ij}=1$. Furthermore, the transition probabilities matrix is defined by
\begin{align*}\label{c03}
\pi=\left[
  \begin{array}{cccc}
   \pi_{11} & \pi_{12} & \cdots & \pi_{1N} \\
    \pi_{21} & \pi_{22} & \cdots & \pi_{2N} \\
     \   & \  & \ddots & \  \\
      \pi_{N1} & \pi_{N2} & \cdots & \pi_{NN} \\
  \end{array}
\right]
\end{align*}
To simplify the notations, let $r_k=i (i\in S)$; thus $A_i$ is an abbreviation for $A(r_k)$, and so on.\\
In addition, the transition probabilities of the Markov chain in this note are considered to be partially available, namely, some elements in matrix $\pi$ are time-invariant but unknown. For instance, a system (1) with four modes will have the transition probabilities matrix $\pi$ as
\begin{eqnarray}\label{c04}
\pi=\left[
  \begin{array}{cccc}
  \pi_{11} & ? & \pi_{13} & ? \\
    ? & ? & ? & \pi_{24} \\
     \pi_{31} & ? & \pi_{33} & ?  \\
      ? & ? & \pi_{43} & \pi_{44} \\
  \end{array}
\right]
\end{eqnarray}
where ''$?$'' represents the unavailable elements. For notation clarity, $\forall i\in S$, we denote
\begin{eqnarray}\label{c05}
S_{\mathcal{K}}^i\triangleq\{j: \text{if}\ \pi_{ij}\ \text{is known} \}, S_{\mathcal{UK}}^i\triangleq\{j: \text{if}\ \pi_{ij}\ \text{is unknown} \}.
\end{eqnarray}
Moreover, if $S_{\mathcal{K}}^i\neq\varnothing$, it is further described as
\begin{eqnarray}\label{c06}
S_{\mathcal{K}}^i=\{\mathcal{K}_1^i,\ldots,\mathcal{K}_s^i \}, 1\leq s\leq N
\end{eqnarray}
where $\mathcal{K}_s^i$ represents the $s$th known element with the index $\mathcal{K}_s^i$ in the $i$th row of matrix $\pi$. For example, in \eqref{c04}, for $i=1$, $S_{\mathcal{K}}^1=\{1,3 \}, S_{\mathcal{UK}}^1=\{2,4 \}$. Also, we denote $\pi_{\mathcal{K}}^i\triangleq\sum_{j\in S_{\mathcal{K}}^i}\pi_{ij}$ throughout the note.
The following assumption is important for obtaining the main results.\\
\begin{remark}\label{cr1}
1
\end{remark}
{\bf Assumption 1}: There exist known constant matrices $F_1$, $F_2$, such that the activation functions $f(x(k))$ satisfies
\begin{displaymath}\label{c07}
[f(x(k))-F_1x(k)]^T[f(x(k))-F_2x(k)]\leq0.
\end{displaymath}
%
%%\label{sec:1}
%%Text with citations \cite{RefB} and \cite{RefJ}.
%%\subsection{Subsection title}
%%\label{sec:2}
%%as required. Don't forget to give each section
%%and subsection a unique label (see Sect.~\ref{sec:1}).
%%\paragraph{Paragraph headings} Use paragraph headings as needed.
%
\subsection{The  Protocol}\label{sec:2.2}
Next, we shall introduce the effects induced by the communication protocol scheduling. In the considered networked system, the Try-Once-Discard protocol is utilized to schedule the signal transmission between sensors and the controller. In networked communication circumstances, communication protocols are always employed to determine which node (or nodes) obtains access to the network at each time instant. The main idea of the protocol scheduling considered in this paper is that only one sensor node is permitted to send data via the communication network at each transmission instant. Let $o(k)\in\{1,2,\ldots,\mathfrak{M} \}$ denote the selected sensor node obtaining access to the communication network at time instant $k$. Then, as shown in [1], due to the scheduling of the WTOD protocol, $o(k)$ can be characterized by the following selection principle:
\begin{equation}\label{c08}
o(k)\triangleq \arg\max_{1\leq m\leq \mathfrak{M}}\|y_m(k)-\bar{y}_m(k-1)\|^2_{Q_m}
\end{equation}
where $\bar{y}_m(k-1)$ represents the previously transmitted signal before time instant $k$ (excluding $k$) associating with the sensor node $m$, and $Q_m(m\in\{1,2,\ldots,\mathfrak{M} \})$ is a known positive definite matrix denoting the weight matrix of the $m$th sensor node.\\
By defining $\bar{Q}\triangleq \text{diag}\{Q_1,Q_2,\ldots,Q_\mathfrak{M}\}$ and $\bar{y}_m(k-1)\triangleq[\bar{y}^T_1(k-1)\ \ \bar{y}^T_2(k-1)\  \cdots\ \bar{y}^T_\mathfrak{M}(k-1) ]^T$, the selection principle \eqref{c08} could be rewritten as follows:
\begin{equation}\label{c09}
o(k)= \arg\max_{1\leq m\leq \mathfrak{M}}\|y(k)-\bar{y}(k-1)\|^2_{\bar{Q}\Phi_m}
\end{equation}
where $\Phi_m=\text{diag}\{\delta(m-1)I,\delta(m-2)I,\ldots,\delta(m-\mathfrak{M})I\},(m\in\{1,2,\ldots,\mathfrak{M} \})$ and $\delta(\cdot)\in\{0,1\}$ is the Kronecker delta function. According to the definition of $\bar{y}(k-1)$, it is easy to see that
\begin{equation}\label{c10}
\bar{y}(k)=\Phi_{o(k)}y(k)+(I-\Phi_{o(k)})\bar{y}(k-1)
\end{equation}
%=
By setting $\bar{x}(k)\triangleq [x^T(k)\ \ \bar{y}^T(k-1)]^T$, under the dispatching of WTOD protocol, network \eqref{c01} is redefined as
\begin{eqnarray}\label{c11}
\left\{
\begin{array}{cll}
 \bar{x}(k+1)&=\bar{A}_{i,o(k)}\bar{x}(k)+\bar{B}_{i}\bar{f}(\bar{x}(k))+\bar{C}_{i}\bar{f}(\bar{x}(k-\tau(k)))+\bar{D}_{1i,o(k)}\bar{\omega}(k)\\
     \bar{y}(k)&=\bar{E}_{i,o(k)}\bar{x}(k)+\bar{D}_{2i,o(k)}\bar{\omega}(k)\\
     \bar{z}(k)&=\bar{M}_{i}\bar{x}(k)\\
\end{array}
\right.
\end{eqnarray}
where
\begin{align*}
&\bar{A}_{i,o(k)}=\left(
   \begin{array}{ccc}
     A_i  & & 0\\
     \Phi_{o(k)} E_i& & I-\Phi_{o(k)}
   \end{array}
 \right),
 \bar{B}_i=\text{diag}\{B_i,\ 0\},
 \bar{C}_i=\text{diag}\{C_i,\ 0\},\bar{M}_i=(M_i\ 0),\\
 &\bar{D}_{1i,o(k)}=\text{diag}\{D_{1i},\ \Phi_{o(k)}D_{2i}\},
\bar{E}_{i,o(k)}=\left(
   \begin{array}{ccc}
     \Phi_{o(k)}E_i  & &  I-\Phi_{o(k)}\\
   \end{array}
 \right),
 \bar{D}_{2i,o(k)}=(0\ \ \Phi_{o(k)}D_{2i}),
\\
 &\bar{f}(x(k))=\mathds{1}_2\otimes f(x(k)),
 \bar{f}(x(k-\tau(k)))=\mathds{1}_2\otimes f(x(k-\tau(k))),\\
& \bar{\omega}(k)=[\omega^T(k)\ \ v^T(k)]^T.
\end{align*}

\subsection{The State Estimator}\label{sec:2.3}
According to the final outputs $\bar{y}(k)$ transmitted through the communication channel with WTOD protocol, the augmented model \eqref{c11} of the state estimator is described by the following expression:
\begin{eqnarray}\label{c12}
\left\{
\begin{array}{cll}
\hat{x}(k+1)&=\bar{A}_{i,o(k)}\hat{x}(k)+\bar{B}_{i}\bar{f}(\hat{x}(k))+\bar{C}_{i}\bar{f}(\hat{x}(k-\tau(k)))+K_{i,o(k)}(\bar{y}(k)-\bar{E}_{i,o(k)}\hat{x}(k))\\
    \hat{z}(k)&=\bar{M}_{i}\hat{x}(k)
\end{array}
\right.
\end{eqnarray}
where $\hat{x}(k)\in \mathbb{R}{^{m+n}}$ is the estimation of $\bar{x}(k)$, $\hat{z}(k)$ is the estimation of the output $z(k)$, and $K_{i,o(k)}$ are the gain matrices to be designed.
%\begin{figure}[h]
%  \centering
%  % Requires \usepackage{graphicx}
%  \includegraphics[width=0.95\textwidth]{fig1}\\
%  \caption{Transmission of measurement signal under the RRP.}\label{fig:1}
%\end{figure}
%\begin{remark}\label{r1}
%To date, several innovative results about the state estimation of MJNNs with delays have been obtained, such as in \cite{Hou-Dong-Wang-Ren,Wu-Su-Chu}. It is important to point out  that these previous work have concluded that the communication protocol must be taken into account. However, in the actual communication channel, it is essential to introduce a certain communication protocol in order to prevent data collisions. Therefore, based on the RRP scheduling, \eqref{a10} formulates a novel state estimator considering the simultaneous effects of mixed time-delays, energy-bounded exogenous disturbances, and the sensor nonlinearities with sector-bounded conditions.
%\end{remark}

Taking $e(k)\triangleq \bar{x}(k)-\hat{x}(k)$ as the estimator error, it is derived from \eqref{c11} and \eqref{c12} that the error of estimation is formulated as
\begin{eqnarray}\label{c13}
\left\{
\begin{array}{cll}
e(k+1)&=\bar{x}(k+1)-\hat{x}(k+1)\\
     &=(\bar{A}_{i,o(k)}-K_{i,o(k)}\bar{E}_{i,o(k)})e(k)+\bar{B}_{i}[\bar{f}(\bar{x}(k))-\bar{f}(\hat{x}(k))]\\
     &\quad+\bar{C}_{i}[\bar{f}(\bar{x}(k-\tau(k)))-\bar{f}(\hat{x}(k-\tau(k)))]+(\bar{D}_{1i,o(k)}-K_{i,o(k)}\bar{D}_{2i,o(k)})\bar\omega(k)\\
     \tilde{z}(k)&=\bar{M}_{i}e(k).
\end{array}
\right.
\end{eqnarray}
By setting $\eta(k)=[\bar{x}^T(k)~\ e^T(k)]^T$ and combining the NNs \eqref{c11} with the error system \eqref{c13}, we derive a compact form for the augmented system:
\begin{eqnarray}\label{c14}
\left\{
\begin{array}{cll}
\eta(k+1)&=\tilde{A}_{i,o(k)}\eta(k)+\tilde{B}_i\tilde{f}(k)+\tilde{C}_i\tilde{f}_{\tau}(k)+\tilde{D}_{i,o(k)}W(k)\\
     \tilde{z}(k)&=\tilde{M}_{i}\eta(k).
\end{array}
\right.
\end{eqnarray}
where
\begin{align*}
&\tilde{f}(k)=[\bar{f}^T(\bar{x}(k))\quad \bar{f}^T(\bar{x}(k))-\bar{f}^T(\hat{x}(k))]^T,\\
&\tilde{f}_{\tau}(k)=[\bar{f}^T(\bar{x}(k-\tau(k)))\quad \bar{f}^T(\bar{x}(k-\tau(k)))-\bar{f}^T(\hat{x}(k-\tau(k)))]^T,\\
&\tilde{A}_{i,o(k)}=\text{diag}\{\bar{A}_{i,o(k)},\ \bar{A}_{i,o(k)}-K_{i,o(k)}\bar{E}_{i,o(k)}\},\tilde{B}_i=\text{diag}\{\bar{B}_i,\ \bar{B}_i\},\\
&\tilde{C}_i=\text{diag}\{\bar{C}_i,\ \bar{C}_i\},\tilde{D}_{i,o(k)}=\text{diag}\{\bar{D}_{1i,o(k)},\ \bar{D}_{1i,o(k)}-K_{i,o(k)}\bar{D}_{2i,o(k)}\},\tilde{M}_i=(0\ \bar{M}_i),\\
&W(k)=[\bar\omega^T(k),\ \bar\omega^T(k)]^T.
\end{align*}
The goal of this paper is to establish a remote state estimator of the form \eqref{c13} such that the following requirements are satisfied simultaneously .

1) The augmented error dynamical system \eqref{c14} with $W(k)\equiv0$ is asymptotically stable in the mean square, if for any initial conditions, the following equality holds:
\begin{equation}\label{c15}
\lim_{k\rightarrow\infty}\mathbb{E}\{\|\eta(k)\|^2\}=0
\end{equation}
2)Under the zero-initial conditions, for a specified disturbance attenuation level $\gamma>0$ and all nonzero $W(k)$, the estimation error $\tilde{z}(k)$ from \eqref{c14} satisfies:
\begin{equation}\label{c16}
\sup_{k}\mathbb{E}\{\|\tilde{z}(k)\|^2\}<\gamma^2\sum_{k=0}^{\infty}\|W(k)\|^2.
\end{equation}
The objective of this paper is to design the state estimator \eqref{c13} such that the augmented error dynamical system \eqref{c14} is asymptotically stable in the mean square with $L_{2}$-$L_{\infty}$ performance $\gamma$.
The  lemmas mentioned below will be necessary to prove our main results.\\
%{\bf Lemma 1 \cite{Liu-Wang-Liang-Liu}}\label{L1}
%Let a positive semi-definite matrix $P\in \mathbb{R}^{n\times n}$ be determined, $y_i\in \mathbb{R}^n$ and $m_i\geq0(i\in \mathbb{Z}^+)$. If the related series satisfy convergence, the following is obtained:
%\begin{eqnarray}\label{a14}
%(\sum_{i=1}^{+\infty}m_iy_i)^TP(\sum_{i=1}^{+\infty}m_iy_i)\leq(\sum_{i=1}^{+\infty}m_i)\sum_{i=1}^{+\infty}m_iy_i^TPy_i.
%\end{eqnarray}
{\bf Lemma 1 \cite{Wang-Wang-Liu10}}\label{C1}
The constant matrices $A_1$,$A_2$,$A_3$ are determined where $A_1=A_1^T$ and $A_2=A_2^T>0$, satisfying $A_1+A_3^TA_2^{-1}A_3<0$ if and only if
\begin{align*}\label{c17}
\left[\begin{array}{cc}
A_1&A_3^T\\
A_3&-A_2
\end{array}\right]
<0
\ \ \text{or}
\left[\begin{array}{cc}
-A_2&A_3\\
A_3^T&A_1
\end{array}\right]
<0.
\end{align*}
{\bf Lemma 2 }\label{C2}
Let $V_0(x),V_1(x),\ldots,V_p(x)$ be quadratic functions of $x\in \mathbb{R}^n$, $V_i(x)=x^TT_ix,\ (i=0,1,\ldots,p)$ with $T_i^T=T_i$. Then, the following is true $V_1(x)\leq0,\ldots,V_p(x)\leq0\Rightarrow V_0(x)\leq0$ if there exist $\gamma_1,\gamma_2,\ldots,\gamma_p>0$ such that
\begin{eqnarray*}\label{c18}
T_0-\sum_{i=1}^{p}\gamma_iT_i\leq0.
\end{eqnarray*}
\section{Main Results}\label{sec:3}
%Here, we first address the dynamical relevant issues and derive a novel stability condition for the augmented system (12). Then, an effective method is given to cope with the design of the estimator by the support of the solvability of LMIs.\\
{\bf Lemma 1}\label{Pro:1}
Let Assumption 1 hold and the gain matrices $K_{i,o(k)}$ of state estimator be determined. Consider the MJNNs \eqref{c01} with completely known transition probabilities \eqref{c05} and time-varying bounded delay under the WTOD protocol, the augmented system \eqref{c14} is asymptotically stable in the mean square if there exist positive scalars $\rho_{1i}$, $\rho_{2i}$, and positive definite matrices $P_{1i,o(k+1)}>0$, $i\in S$, $Z>0$ such that
\begin{eqnarray}\label{c19}
\left(
  \begin{array}{cc}
    -\bar{P}_{i,o(k+1)} &\Omega_{1i,o(k)} \\
   * & \Omega_{2i,o(k)}\\
  \end{array}
\right)
<0.
\end{eqnarray}
where
\begin{align*}
&\Omega_{1i}=[\bar{P}_{i,o(k+1)}\tilde{A}_{i,o(k)}\ \ 0\ \ \bar{P}_{i,o(k+1)}\tilde{B}_i\ \ \bar{P}_{i,o(k+1)}\tilde{C}_i],\\
&\Omega_{2i}=\left(
           \begin{array}{cccc}
             \Omega_{11} & 0 & \rho_{1i}F_4 & 0 \\
             * & -\rho_{2i}F_3-Z & 0 & \rho_{2i}F_4 \\
             * & * & -\rho_{1i}I & 0 \\
             * & * & * & -\rho_{2i}I
           \end{array}
         \right),\\
&\Omega_{11}=-P_{i,o(k)}+(1+\tau-\bar{\tau})Z-\rho_{1i}F_3-\check{E}_{i}^T\nu\check{E}_{i},\\
&\nu=\sum_{m=1}^\mathfrak{M}\sigma_{m}(k)\bar{Q}(\Phi_m-\Phi_{o(k)}),P_{i,o(k)}=\text{diag}\{P_{1i,o(k)},P_{1i,o(k)} \},\\
&\bar{P}_{i,o(k+1)}=\text{diag}\{\bar{P}_{1i,o(k+1)},\bar{P}_{1i,o(k+1)} \},\bar{P}_{1i,o(k+1)}\triangleq\sum_{j\in S}\pi_{ij}P_{1j,o(k+1)},\\
&F_3=\frac{I\otimes(F_1^TF_2+F_1F_2^T)}{2},F_4 =\frac{(I\otimes(F_1+F_2))^T}{2}.
\end{align*}
{\bf Proof}: The Lyapunov-Krasovskii functional is selected as follows
\begin{equation}\label{c20}
V(k)=V_1(\eta(k),r_k)+V_2(\eta(k),r_k)
\end{equation}
in which
\begin{align*}
&V_1(\eta(k),r_k)=\eta^T(k)P_{i,o(k)}\eta(k), \\
&V_2(\eta(k),r_k)=\sum_{d=k-\tau(k)}^{k-1}\eta^T(d)Z\eta(d)+\sum_{l=k-\tau+1}^{k-\bar{\tau}}\sum_{d=l}^{k-1}\eta^T(d)Z\eta(d).
\end{align*}
For $i\in S$, we have
\begin{align}\label{c21}
&\mathbb{E}\{\bigtriangleup V_1(\eta(k),r_k)|W(k)=0\}\nonumber\\
=&\mathbb{E}\{[\tilde{A}_{i,o(k)}\eta(k)+\tilde{B}_i\tilde{f}(k)+\tilde{C}_i\tilde{f}_{\tau}(k)]^T\bar{P}_{i,o(k+1)}[\tilde{A}_{i,o(k)}\eta(k)+\tilde{B}_i\tilde{f}(k)+\tilde{C}_i\tilde{f}_{\tau}(k)]\nonumber\\
&-\eta^T(k)P_{i,o(k)}\eta(k)\}\nonumber\\
\end{align}
and
\begin{align}\label{c22}
&\mathbb{E}\{\bigtriangleup V_2(\eta(k),r_k)|W(k)=0\}\nonumber\\
=&\mathbb{E}\{\sum_{d=k+1-\tau(k+1)}^{k}\eta^T(d)Z\eta(d)+\sum_{l=k+1-\tau+1}^{k+1-\bar{\tau}}\sum_{d=l}^{k}\eta^T(d)Z\eta(d)\nonumber\\
&-\sum_{d=k-\tau(k)}^{k-1}\eta^T(d)Z\eta(d)-\sum_{l=k-\tau+1}^{k-\bar{\tau}}\sum_{d=l}^{k-1}\eta^T(d)Z\eta(d)\}\nonumber\\
=&\mathbb{E}\{\sum_{d=k+1-\tau(k+1)}^{k-\bar{\tau}}\eta^T(d)Z\eta(d)+\sum_{d=k+1-\bar{\tau}}^{k-1}\eta^T(d)Z\eta(d)+\eta^T(k)Z\eta(k)\nonumber\\
&-\sum_{d=k+1-\tau(k)}^{k-1}\eta^T(d)Z\eta(d)-\eta_{\tau}^T(k)Z\eta_{\tau}(k)+(\tau-\bar{\tau})\eta^T(k)Z\eta(k)\nonumber\\
&-\sum_{l=k+1-\tau}^{k-\bar{\tau}}\eta^T(l)Z\eta(l))\}\nonumber\\
\leq&\mathbb{E}\{(1+\tau-\bar{\tau})\eta^T(k)Z\eta(k)-\eta_{\tau}^T(k)Z\eta_{\tau}(k)\}.
\end{align}
where $\eta_{\tau}(k)=\eta(k-\tau(k))$.
It is readily observed from \eqref{c21}-\eqref{c22} that
\begin{align}\label{c23}
&\mathbb{E}\{\bigtriangleup V(\eta(k),r_k)|W(k)=0\}\nonumber\\
\leq&\mathbb{E}\{[\tilde{A}_{i,o(k)}\eta(k)+\tilde{B}_i\tilde{f}(k)+\tilde{C}_i\tilde{f}_{\tau}(k)]^T\bar{P}_{i,o(k+1)}[\tilde{A}_{i,o(k)}\eta(k)+\tilde{B}_i\tilde{f}(k)+\tilde{C}_i\tilde{f}_{\tau}(k)]\nonumber\\
&-\eta^T(k)P_{i,o(k)}\eta(k)+(1+\tau-\bar{\tau})\eta^T(k)Z\eta(k)-\eta_{\tau}^T(k)Z\eta_{\tau}(k)\}.
\end{align}
According to Assumption 1, we obtain that
\begin{eqnarray}\label{c24}
&\rho_{_{1i}}\left[
  \begin{array}{c}
  \eta(k)  \\
   \tilde{f}(k)
  \end{array}
\right]^T
\left[
  \begin{array}{cc}
  F_3 & -F_4\\
  * & I
  \end{array}
\right]
\left[
  \begin{array}{c}
  \eta(k)  \\
   \tilde{f}(k)
  \end{array}
\right]
\leq0
\end{eqnarray}
\begin{eqnarray}\label{c25}
&\rho_{_{2i}}\left[
  \begin{array}{c}
  \eta_{\tau}(k)  \\
   \tilde{f}_{\tau}(k)
  \end{array}
\right]^T
\left[
  \begin{array}{cc}
  F_3 & -F_4\\
  * & I
  \end{array}
\right]
\left[
  \begin{array}{c}
  \eta_{\tau}(k)  \\
   \tilde{f}_{\tau}(k)
  \end{array}
\right]
\leq0
\end{eqnarray}
By analyzing the scheduling mechanism of WTOD protocol \eqref{c09}, we obtain that, for any $m\in \mathfrak{M}$
\begin{align*}\label{c26}
\mathbb{E}\{[y(k)-\bar{y}(k-1)]^T\bar{Q}(\Phi_m-\Phi_{o(k)})[y(k)-\bar{y}(k-1)] |W(k)=0\}\leq0
\end{align*}
which can be written in terms of $\eta(k)$ as
\begin{equation*}\label{c27}
\mathbb{E}\{\eta^T(k)\check{E}_{i}^T\bar{Q}(\Phi_m-\Phi_{o(k)})\check{E}_{i}\eta(k) |W(k)=0\}\leq0
\end{equation*}
where $\check{E}_{i}=[E_{i}\ -I\ 0]$.\\
%\begin{equation*}\label{c27}
%\mathbb{E}\{\zeta^T(k)\Psi_{i}^T\bar{Q}(\Phi_m-\Phi_{o(k)})\Psi_{i}\varsigma(k) \}\leq0
%\end{equation*}
%where $\zeta^T(k)=[\eta^T(k)\ W^T(k)]^T, \Psi_i=[\check{E}_{i}\ 0\ 0\ \check{D}_{i}], \check{E}_{i}=[\tilde{E}_{i}\ -I\ 0], \check{D}_{i}=[0\ D_{2i}]$.
According to Lemma 2, if there exist $\sigma_1(k),\sigma_2(k),\ldots,\sigma_\mathfrak{M}(k)>0$ such that
\begin{align}\label{c28}
-\mathbb{E}\{\eta^T(k)\check{E}_{i}^T\sum_{m=1}^\mathfrak{M}\sigma_{m}(k)\bar{Q}(\Phi_m-\Phi_{o(k)})\check{E}_{i}\eta(k) \}\geq0
\end{align}
By substituting \eqref{c24}-\eqref{c28} into \eqref{c23}, we obtain
\begin{align}\label{a29}
&\mathbb{E}\{\bigtriangleup V(\eta(k),r_k)|W(k)=0\}\nonumber\\
\leq&\mathbb{E}\{[\tilde{A}_{i,o(k)}\eta(k)+\tilde{B}_i\tilde{f}(k)+\tilde{C}_i\tilde{f}_{\tau}(k)]^T\bar{P}_{i,o(k+1)}[\tilde{A}_{i,o(k)}\eta(k)+\tilde{B}_i\tilde{f}(k)+\tilde{C}_i\tilde{f}_{\tau}(k)]\nonumber\\
&-\eta^T(k)P_{i,o(k)}\eta(k)+(1+\tau-\bar{\tau})\eta^T(k)Z\eta(k)-\eta_{\tau}^T(k)Z\eta_{\tau}(k)\nonumber\\
&-\eta^T(k)\check{E}_{i}^T\nu\check{E}_{i}\eta(k)
-\rho_{_{1i}}\left[
  \begin{array}{c}
  \eta(k)  \\
   \tilde{f}(k)
  \end{array}\right]^T
\left[
  \begin{array}{cc}
  F_3 & -F_4\\
  * & I
  \end{array}\right]
\left[
  \begin{array}{c}
  \eta(k)  \\
   \tilde{f}(k)
  \end{array}\right]\nonumber\\
&-\rho_{_{2i}}\left[
  \begin{array}{c}
  \eta_{\tau}(k)  \\
   \tilde{f}_{\tau}(k)
  \end{array}\right]^T
\left[ \begin{array}{cc}
  F_3 & -F_4\\
  * & I
  \end{array}\right]
\left[
  \begin{array}{c}
  \eta_{\tau}(k)  \\
   \tilde{f}_{\tau}(k)
  \end{array}\right]  \}\nonumber\\
\leq&\xi^T(k)\Omega_{i,o(k)}\xi(k)
\end{align}
in which
\begin{eqnarray*}
&\xi^T(k)=[\eta^T(k)\ \ \eta^T_{\tau}(k)\ \ \tilde{f}^T(k)\ \ \tilde{f}_{\tau}^T(k)]^T\\
&\Omega_{i,o(k)}=\bar{\Omega}^T_{1i,o(k)}\bar{P}_{i,o(k+1)}\bar{\Omega}_{1i,o(k)}+\Omega_{2i,o(k)}, \bar{\Omega}_{1i,o(k)}=[\tilde{A}_{i,o(k)}\ \ 0\ \ \tilde{B}_i\ \ \tilde{C}_i]
\end{eqnarray*}
Letting $\varpi=\lambda_{\max}(\Omega_{i,o(k)})$, we obtain that
\begin{align}\label{c30}
\mathbb{E}\{\bigtriangleup V(\eta(k),r_k)|W(k)=0\}\leq\varpi\mathbb{E}\{\|\xi(k)\|^2 \}.
\end{align}
Summing both sides of \eqref{c30} from $0$ to $N$ regarding $k$ leads to
\begin{align}\label{c31}
\mathbb{E}\{ V(\eta(N+1),r_{N+1})|W(k)=0\}-\mathbb{E}\{V(\eta(0),r_0)|W(k)=0\}\leq\varpi\sum_{k=0}^N\mathbb{E}\{\|\xi(k)\|^2 \}
\end{align}
which further indicates
\begin{align*}\label{c32}
\sum_{k=0}^N\mathbb{E}\{\|\xi(k)\|^2 \}\leq-\frac{1}{\varpi}\mathbb{E}\{V(\eta(0),r_0)|W(k)=0\}.
\end{align*}
We can draw the conclusion that the series $\sum_{k=0}^N\mathbb{E}\{\|\xi(k)\|^2 \}$ is convergent, and hence
\begin{equation*}\label{c33}
\lim_{k\rightarrow\infty}\mathbb{E}\{\|\xi(k)\|^2 \}=0,
\end{equation*}
which implies that the system \eqref{c14} is asymptotically stable in the mean square and the proof is now complete.\\
\begin{remark}\label{rrr1}
With partially known transition probabilities, $\bar{P}_{1i,,o(k+1)}=\sum_{j\in S_{\mathcal{K}}^i}\pi_{ij}P_{1j,o(k+1)}+(1-\pi_i^{k})\sum_{j\in S_{\mathcal{UK}}^i}P_{1j,o(k+1)}$. When $\bar{P}_{1i,o(k+1)}=\sum_{j\in S_{\mathcal{K}}^i}\pi_{ij}P_{1j,o(k+1)}$ in Theorem 1, the conditions reduce to asymptotically stable with completely known transition probabilities, that is, $S_{\mathcal{K}}^i=S$ and $S_{\mathcal{UK}}^i=\varnothing$. When  $\bar{P}_{1i,o(k+1)}=P_{1j,o(k+1)}$ in Theorem 1, the conditions are reduced to asymptotically stable with completely unknown transition probabilities. That is, both asymptotically stable with completely known transition probabilities or with completely unknown transition probabilities can be seen as special cases of the considered case.
\end{remark}
In the following Theorem, a sufficient condition is obtained that guarantees the augmented error system \eqref{c14} asymptotically stable in the mean square.\\
{\bf Theorem 1}\label{Th:1}
Let Assumption 1 hold and the gain matrices $K_{i,o(k)}$ of state estimator be determined. Consider the MJNNs \eqref{c01} with partially known transition probabilities \eqref{c05} and time-varying bounded delay under the WTOD protocol. the augmented system \eqref{c14} is asymptotically stable in the mean square if there exist positive scalars $\rho_{1i}$, $\rho_{2i}$, and positive definite matrices $P_{1i,o(k+1)}>0$, $i\in S$, $Z>0$ such that
\begin{eqnarray}\label{c34}
\left(
  \begin{array}{cc}
    -\Upsilon_{j} &\Gamma_{1i,o(k)} \\
   * & \Omega_{2i,o(k)}\\
  \end{array}
\right)
<0
\end{eqnarray}
where $\Gamma_{1i,o(k)}=[\Upsilon_{j}\tilde{A}_{i,o(k)}\ \ 0\ \ \Upsilon_{j}\tilde{B}_i\ \ \Upsilon_{j}\tilde{C}_i]$ and $\Omega_{2i,o(k)}$ is defined in Proposition 1 and if $\pi_{\mathcal{K}}^i=0$, $\Upsilon_{j}\triangleq P_{j}$, otherwise,
\begin{eqnarray*}\label{c35}
\left\{
\begin{array}{cll}
\Upsilon_{j}&\triangleq \frac{1}{\pi_{\mathcal{K}}^i}P_{\mathcal{K}}^i\\
\Upsilon_{j}&\triangleq P_{j}, \forall j\in S_{\mathcal{UK}}^{i}\\
\end{array}
\right.
\end{eqnarray*}
with
\begin{eqnarray*}
P_{\mathcal{K}}^i=\sum_{j\in S_{\mathcal{K}}^i}\pi_{ij}
\left(
  \begin{array}{cc}
    -P_{1j,o(k+1)} & 0 \\
   * & P_{1j,o(k+1)}\\
  \end{array}
\right),
P_{j}=
\left(
  \begin{array}{cc}
    -P_{1j,o(k+1)} & 0 \\
   * & P_{1j,o(k+1)}\\
  \end{array}
\right).
\end{eqnarray*}
%$P_{\mathcal{K}}^i\triangleq\sum_{j\in S_{\mathcal{K}}^i}\pi_{ij}P_{j}$.\\
{\bf Proof}: First of all, we know that the augmented error system \eqref{c14} is asymptotically stable under the completely known transition probabilities \eqref{c05} if \eqref{c19} holds. Note that \eqref{c19} can be rewritten as
\begin{eqnarray*}\label{c36}
\Xi_i=
\left[
  \begin{array}{cc}
    -P_{\mathcal{K}}^i & P_{\mathcal{K}}^i\bar{\Omega}_{1i,o(k)} \\
   * & \pi_{\mathcal{K}}^i\Omega_{2i,o(k)}\\
  \end{array}
\right]
+
\sum_{j\in S_{\mathcal{UK}}^i}\pi_{ij}
\left[
  \begin{array}{cc}
    -P_{j} & P_{j}\bar{\Omega}_{1i,o(k)} \\
   * & \Omega_{2i,o(k)}\\
  \end{array}
\right]
\end{eqnarray*}
Therefore, if one has
\begin{eqnarray}\label{c37}
\left[
  \begin{array}{cc}
    -P_{\mathcal{K}}^i & P_{\mathcal{K}}^i\bar{\Omega}_{1i,o(k)} \\
   * & \pi_{\mathcal{K}}^i\Omega_{2i,o(k)}\\
  \end{array}
\right]
<0
\end{eqnarray}
\begin{eqnarray}\label{c38}
\left[
  \begin{array}{cc}
    -P_{j} & P_{j}\bar{\Omega}_{1i,o(k)} \\
   * & \Omega_{2i,o(k)}\\
  \end{array}
\right]
<0,
\forall j\in S_{\mathcal{UK}}^i,
\end{eqnarray}
then we have $\Xi_i<0$, hence the system \eqref{c14} is asymptotically stable under partially known transition probabilities, which is concluded from the obvious fact that no knowledge on $\pi_{ij}$, $n\in S_{\mathcal{UK}}^i$ is required in \eqref{c37} and \eqref{c38}. Thus, for $\pi_{\mathcal{K}}^i\neq0$ and $\pi_{\mathcal{K}}^i=0$, respectively, one can readily obtain \eqref{c36}, since if $\pi_{\mathcal{K}}^i=0$, the conditions \eqref{c37}, \eqref{c38} will reduce to \eqref{c38}. This completes the proof.\\
In the following proposition, a sufficient condition is obtained that guarantees the augmented error system \eqref{c14} asymptotically stable in the mean square with $L_2-L_\infty$ performance $\gamma$.\\
Next, we consider the augmented system \eqref{c14} is asymptotically stable with $L_2-L_\infty$ performance $\gamma$.\\
{\bf Theorem 2}\label{Th:2}
Under Assumption 1, for given scalar $\gamma$, the estimator gain matrices $K_{i,o(k)}$, the augmented system \eqref{c14} with partially known transition probabilities \eqref{c05} and time-varying bounded delay under the WTOD protocol is asymptotically stable in the mean square with $L_{2}$-$L_{\infty}$ performance $\gamma$ if there exist positive scalars $\rho_{1i}$, $\rho_{2i}$, and positive definite matrices $P_{1i,o(k+1)}>0$, $i\in S$, $Z>0$ such that
\begin{eqnarray}\label{c39}
\left(
  \begin{array}{cc}
    -\Upsilon_{j} &\bar{\Gamma}_{1i,o(k)} \\
   * & \tilde{\Omega}_{2i,o(k)}\\
  \end{array}
\right)
<0,
\end{eqnarray}
\begin{eqnarray}\label{c40}
\left(
  \begin{array}{cc}
    P_{i,o(k)} &\tilde{M}^T_{i} \\
   * & \gamma^2I\\
  \end{array}
\right)
>0
\end{eqnarray}
where
\begin{align*}
&\bar{\Gamma}_{1i,o(k)}=[\Upsilon_{j}\tilde{A}_{i,o(k)}\ \ 0\ \ \Upsilon_{j}\tilde{B}_i\ \ \Upsilon_{j}\tilde{C}_i\ \ \Upsilon_{j}\tilde{D}_{i,o(k)}],\\
&\tilde{\Omega}_{2i,o(k)}=\left(
           \begin{array}{cc}
             \Omega_{2i,o(k)} & \check{E}_{i}^T\nu \hat{D}_{i} \\
              * &     -I-\hat{D}_{i}^T\nu \hat{D}_{i}\\
           \end{array}
         \right).
\end{align*}
{\bf Proof}: In order to discuss the  $L_{2}$-$L_{\infty}$ disturbance attenuation level of the estimator, the same Lyapunov-Krasovskii is chosen as that in proof of Proposition 1 with $W(k)\neq0$, under scheduling of WTOD protocol we have
\begin{equation*}
\mathbb{E}\{\bar{\xi}^T(k)\Psi_{i}^T\sum_{m=1}^\mathfrak{M}\sigma_{m}(k)\bar{Q}(\Phi_m-\Phi_{o(k)})\Psi_{i}\bar{\xi}(k) \}\leq0
\end{equation*}
where
\begin{align*}
&\bar{\xi}(k)\triangleq[\xi^T(k)\ \ W^T(k)]^T, \Psi_i=[\check{E}_{i}\ \hat{D}_{i}],\\
&\check{E}_{i}=[E_{i}\ -I\ 0], \hat{D}_{i}=[0\ 0\ \check{D}_{i}], \check{D}_{i}=[0\ D_{2i}]
\end{align*}
Consequently, a similar derivation yields
\begin{align}\label{c41}
\mathbb{E}\{\bigtriangleup V(\eta(k),r_k)\}\leq\bar{\xi}^T(k)\bar{\Omega}_{i,o(k)}\bar{\xi}(k)
\end{align}
where
\begin{align*}
&\bar{\Omega}_{i,o(k)}=\tilde{\Omega}^T_{1i,o(k)}\bar{P}_{i,o(k+1)}\tilde{\Omega}_{1i,o(k)}+\bar{\Omega}_{2i,o(k)},\\
&\tilde{\Omega}_{1i,o(k)}=[\tilde{A}_{i,o(k)}\ \ 0\ \ \tilde{B}_i\ \ \tilde{C}_i\ \ \tilde{D}_{i,o(k)}],
\bar{\Omega}_{2i,o(k)}=\left(
           \begin{array}{cc}
             \Omega_{2i,o(k)} & \check{E}_i^T\nu \hat{D}_i \\
              * &  -\hat{D}_i^T\nu \hat{D}_i\\
           \end{array}
         \right).
\end{align*}
By employing Schur complement to \eqref{c41}, it can be easily seen that
\begin{align*}
\mathbb{E}\{\bigtriangleup V(\eta(k),r_k)\}\leq\bar{\xi}^T(k)\left(
  \begin{array}{cc}
    -\bar{P}_{i,o(k+1)} &\tilde{\Omega}_{1i,o(k)} \\
   * & \bar{\Omega}_{2i,o(k)}\\
  \end{array}
\right)\bar{\xi}(k)
\end{align*}
It is implies from \eqref{c39} that $\mathbb{E}\{\bigtriangleup V(\eta(k),r_k)\}\leq0$.\\
Under the zero-initial conditions, the following cost function is constructed:
\begin{equation}\label{c42}
\sup_{k}\mathbb{E}\{\|\tilde{z}(k)\|^2\}-\gamma^2\sum_{k=0}^{\infty}\|W(k)\|^2.
\end{equation}
Note that
\begin{eqnarray}\label{c43}
\begin{array}{ll}
&\mathbb{E}\{V(\eta(k),r_k)\}-\sum_{l=0}^{k-1}W^T(l)W(l)\nonumber\\
=&\sum_{l=0}^{k-1}(\mathbb{E}\{\bigtriangleup V(\eta(l),r_l)\}-W^T(l)W(l))\nonumber\\
\leq&\sum_{l=0}^{k-1}(\bar{\xi}^T(l)\bar{\Omega}_{i,o(k)}\bar{\xi}(l)-W^T(l)W(l))\nonumber\\
=& \sum_{l=0}^{k-1}\bar{\xi}^T(l)\tilde{\Omega}_{i,o(k)}\bar{\xi}(l).
\end{array}
\end{eqnarray}

The inequality $\tilde{\Omega}_{i,o(k)}<0$ from \eqref{c39} tells
\begin{eqnarray}\label{c44}
\mathbb{E}\{V(\eta(k),r_k)\}<\sum_{l=0}^{k-1}W^T(l)W(l)
\end{eqnarray}
By employing Schur complement to \eqref{c40}, it can be easily seen that
\begin{eqnarray}\label{c45}
\tilde{M}^T_{i}\tilde{M}_{i}<\gamma^2P_{i,o(k)}
\end{eqnarray}
Taking \eqref{c43}-\eqref{c45} into consideration, one has
\begin{equation}\label{c46}
\begin{array}{rl}
\mathbb{E}\{\tilde{z}^T(k)\tilde{z}(k)\}&=\mathbb{E}\{\eta^T(k)\tilde{M}^T_{i}\tilde{M}_{i}\eta(k)\}\nonumber\\
&\leq \gamma^2\mathbb{E}\{\eta^T(k)P_{i,o(k)}\eta(k)\}\nonumber\\
&< \gamma^2\mathbb{E}\{V(\eta(k),r_k)\}\nonumber\\
&\leq \gamma^2\sum_{l=0}^{k-1}W^T(l)W(l).
\end{array}
\end{equation}
Taking the supremum of $\mathbb{E}\{\tilde{z}^T(k)\tilde{z}(k)\}$ over $k$ and the limit of $\sum_{l=0}^{k-1}W^T(l)W(l)$ with $k\rightarrow\infty$, we have
\begin{equation}\label{c47}
\sup_{k}\mathbb{E}\{\|\tilde{z}(k)\|^2\}<\gamma^2\sum_{k=0}^{\infty}\|W(k)\|^2.
\end{equation}
Hence, condition \eqref{c16} is fulfilled under the zero initial conditions for any non-zero $W(k)$. This completes the proof.\\
Now, the following theorem presents a sufficient condition for the asymptotical stability of system \eqref{c14} with partially known transition probabilities \eqref{c05}.\\
Now let us consider the stabilizing controller design. From the above development, it can be seen that the system with completely known transition probabilities is just a special case of our considered systems. In what follows, we will give a stabilization condition of the system with partially known transition probabilities as generalized results.\\
{\bf Theorem 3}\label{Th:3}
Under Assumption 1,  consider the augmented system \eqref{c14} with partially known transition probabilities \eqref{c05} under the WTOD protocol is asymptotically stable in the mean square with $L_{2}$-$L_{\infty}$ performance $\gamma$ if there exist positive scalars $\rho_{1i}$, $\rho_{2i}$, and positive definite matrices $P_{1i,o(k+1)}>0$, $X_{i,o(k)}$,  $i\in S$, $Z>0$, and $K_{i,o(k)}$ such that
\begin{eqnarray}\label{c48}
\left(
  \begin{array}{cc}
    -\hat{\Upsilon}_{j} & L_{\mathcal{K}}^i\tilde{\Omega}_{1i,o(k)} \\
   * & \tilde{\Omega}_{2i,o(k)}\\
  \end{array}
\right)
<0
\end{eqnarray}
\begin{eqnarray}\label{c49}
\left(
  \begin{array}{cc}
    P_{i,o(k)} &\tilde{M}^T_{i} \\
   * & \gamma^2I\\
  \end{array}
\right)
>0
\end{eqnarray}
\begin{equation}\label{c50}
P_{i,o(k+1)}X_{i,o(k+1)}=I
\end{equation}
where $\tilde{\Omega}_{1i,o(k)}$, $\tilde{\Omega}_{2i,o(k)}$ is defined in Theorem 2 and if $\pi_{\mathcal{K}}^i=0$, $\hat{\Upsilon}_{j}\triangleq X_{j}$ and $L_{\mathcal{K}}^i\triangleq I$, otherwise,
\begin{eqnarray}\label{c51}
\left\{
\begin{array}{cll}
\hat{\Upsilon}_{j}=&\pi_{\mathcal{K}}^i\text{diag}\{X_{\mathcal{K}_1^i},X_{\mathcal{K}_2^i},\cdots,X_{\mathcal{K}_s^i} \}\\
L_{j}=&[\sqrt{\pi_{i\mathcal{K}_1^i}}I,\cdots,\sqrt{\pi_{i\mathcal{K}_s^i}}I]^T\\
\hat{\Upsilon}_{j}=&X_{j}, L_{j}=I \  \forall j\in S_{\mathcal{UK}}^{i}\\
\end{array}
\right.
\end{eqnarray}

%with $P_{\mathcal{K}}^m\triangleq\sum_{n\in S_{\mathcal{K}}^m}\pi_{mn}P_{n}$.\\
{\bf Proof}: By Schur complement, \eqref{c34} is equivalent to (for $\pi_{\mathcal{K}}^i\neq0$ )
\begin{eqnarray}\label{c52}
\left[
  \begin{array}{cc}
    \Xi_{3i} & \Xi_{4i}\tilde{\Omega}_{1i,o(k)} \\
   * & \tilde{\Omega}_{2i,o(k)}\\
  \end{array}
\right]
<0
\end{eqnarray}
\begin{eqnarray}\label{c53}
\left[
  \begin{array}{cc}
    -P^{-1}_{j} & \tilde{\Omega}_{1i,o(k)} \\
   * & \Omega_{2i,o(k)}\\
  \end{array}
\right]
<0,
\forall j\in S_{\mathcal{UK}}^j,
\end{eqnarray}
where{}
\begin{eqnarray*}
&\Xi_{3i}=\left[
  \begin{array}{cccc}
    -\pi_{\mathcal{K}}^iP_{\mathcal{K}_1^i}^{-1}  & 0 & \cdots  &  0 \\
   * &  \pi_{\mathcal{K}}^iP_{\mathcal{K}_2^i}^{-1}  &  \cdots & 0\\
   * & *  &  \ddots  & \vdots\\
   * & *  & *  &  -\pi_{\mathcal{K}}^iP_{\mathcal{K}_s^i}^{-1} \\
  \end{array}
\right]\\
&\Xi_{4i}=[\sqrt{\pi_{i\mathcal{K}_1^i}}I,\cdots,\sqrt{\pi_{i\mathcal{K}_s^i}}I]^T
\end{eqnarray*}
Note that if $\pi_{\mathcal{K}}^i=0$, \eqref{c34} will be just equivalent to  \eqref{c53}. Setting $X_{i,o(k+1)}=P_{i,o(k+1)}^{-1}$, $\hat{\Upsilon}_{j}$ and $L_{\mathcal{K}}^i$ as shown in \eqref{c51}, we can readily obtain \eqref{c48} and \eqref{c50}. This completes the proof.\\
It should be noted that the criteria in Theorems 3 are not strict linear matrix inequalities because the existence $P_{i,o(k+1)}X_{i,o(k+1)}=I$, which can be solved by using the cone complementarity linearization method [1]. In the following, an algorithm is proposed for Theorem 3.\\
{\bf Algorithm 1.}\label{Al:1} Given constants $d,$ and let $c$ denotes the maximum number of iterations.\\
\begin{enumerate}
  \item Find a flexible solution $\{P_{i,o(k+1)}, X_{i,o(k+1)}\}$ to LMIs \eqref{c48} and \eqref{c49}
   \begin{eqnarray}\label{c54}
\left(
  \begin{array}{cc}
    P_{i,o(k+1)} &I \\
   * & X_{i,o(k+1)}\\
  \end{array}
\right)
\geq0, i=1,2,\ldots,N.\\
\end{eqnarray}
If no feasible solution, EXIT. Else, set $t=0$.
  \item Solve the following minimization problem:
  $\min tr(\sum_{i=1}^{N}(P_{it,o(k+1)}X_{i,o(k+1)}+X_{it,o(k+1)}P_{i,o(k+1)}))$ subject to LMIs \eqref{c48}, \eqref{c49}and \eqref{c54}.
  \item If \eqref{c55} is satisfied for a sufficient small scalar $\mu>0$, output the feedback gain $K_{i,o(k)}$.\\
  Otherwise, set $t=t+1$. If $t<c$ (c denotes the maximum number of iterations), go to Step 2, otherwise, EXIT.
     \begin{eqnarray}\label{c55}
|tr(\sum_{i=1}^{N}(P_{it,o(k+1)}X_{i,o(k+1)}+X_{it,o(k+1)}P_{i,o(k+1)}))-2Nn|<\mu,
\end{eqnarray}
where $n$ is the dimension of $P_{i,o(k+1)}$.
\end{enumerate}

%\begin{remark}\label{r2}
%Within the RRP scheduling, Theorem 1 makes a significant contribution to the asymptotical stability of estimation error dynamics that show the prescribed $H_\infty$ performance, which is quite different from the existing results for the MJNNs on the exponential state estimation problem reported in \cite{Li-Dong-Wang-Zhang,Zhang-Yu}. Making use of a suitable Lyapunov-Krasovskii functional and the stochastic mathematical approaches, the asymptotical stability can be eventually  demonstrated by finding a series of LMIs solutions, and implies  that the stability criterion depends strongly on both the communication protocol and the Markovian switching signal.
%\end{remark}
%\begin{remark}
%With regard to Theorem 2, gain matrices are carefully solved so as to eventually complete the design of the estimator influenced by both the RRP and the Markovian switching.
%Different from the results reported in \cite{Wan-Wang-Wu-Liu},  the estimator given in Theorem 2 can deal with the effects resulting from mixed time-delays and the sensors nonlinearities making our result generally applicable in practice.
%\end{remark}

\section{A Numerical Example}\label{sec:4}
Here, a numerical example is used, and the simulations  for the example confirm the validity of the theoretical conclusions.

Let system \eqref{c01} be two-neuron and four-mode neural network parameters as follows:
\begin{align*}
%&\tau(1)=1, \tau(2)=5, \underline{\tau}=1,\bar{\tau}=5,\gamma=1.9246, v_l=2^{-l},\\
&A(1)=\left(
   \begin{array}{cc}
   0.27 & 0 \\
   0 & 0.63 \\
   \end{array}
 \right),
A(2)=\left(
   \begin{array}{cc}
   0.32 & 0 \\
   0.16 & 0.47 \\
   \end{array}
 \right),\\
 &A(3)=\left(
   \begin{array}{cc}
   0.30 & 0.13 \\
   0.16 & 0.14 \\
   \end{array}
 \right),
A(4)=\left(
   \begin{array}{cc}
   0.50 & 0 \\
   0.21 & 0.29 \\
   \end{array}
 \right),\\
 &B(1)=\left(
   \begin{array}{cc}
   -0.50 & 0.80 \\
   0.30 & -0.33 \\
   \end{array}
 \right),
B(2)=\left(
   \begin{array}{cc}
   0.20 & -0.70 \\
   -0.55 & 0.62 \\
   \end{array}
 \right),\\
 &B(3)=\left(
   \begin{array}{cc}
   -0.80 & 0.70 \\
   0.50 & 0.38 \\
   \end{array}
 \right),
B(4)=\left(
   \begin{array}{cc}
   0.40 & -0.60 \\
   -0.40 & 0.60 \\
   \end{array}
 \right),\\
 &C(1)=\left(
   \begin{array}{cc}
   -0.02 & 0.12 \\
   0.07 & -0.14 \\
   \end{array}
 \right),
C(2)=\left(
   \begin{array}{cc}
   0.02 & 0.12 \\
   0.07 & 0.02 \\
   \end{array}
 \right),\\
 &C(3)=\left(
   \begin{array}{cc}
   -0.02 & 0.12 \\
   0.07 & 0.02 \\
   \end{array}
 \right),
C(4)=\left(
   \begin{array}{cc}
   0.02 & 0.12 \\
   0.07 & -0.14 \\
   \end{array}
 \right),\\
 &D(11)=\text{diag}\{-0.04,-0.04\},D(12)=\text{diag}\{-0.03,-0.04\},\\
 &D(13)=\text{diag}\{-0.02,-0.03\},D(14)=\text{diag}\{-0.05,-0.04\},\\
&D(21)=\left(
   \begin{array}{cc}
   -0.11 & 0.15 \\
   0.12 & 0.16 \\
   \end{array}
 \right),
D(22)=\left(
   \begin{array}{cc}
   -0.20 & 0.18 \\
   0.10 & 0.06 \\
   \end{array}
 \right),\\
 &D(23)=\left(
   \begin{array}{cc}
   -0.21 & 0.05 \\
   0.11 & 0.15 \\
   \end{array}
 \right),
D(24)=\left(
   \begin{array}{cc}
   -0.12 & 0.14 \\
   0.20 & 0.12 \\
   \end{array}
 \right),\\
  &E(1)=\left(
   \begin{array}{cc}
   0.10 & 0.20 \\
   0.15 & -0.20 \\
   \end{array}
 \right),
E(2)=\left(
   \begin{array}{cc}
   -0.25 & 0.15 \\
   0.15 & 0.20 \\
   \end{array}
 \right),\\
 &E(3)=\left(
   \begin{array}{cc}
   -0.20 & 0.15 \\
   0.15 & -0.10 \\
   \end{array}
 \right),
E(4)=\left(
   \begin{array}{cc}
   -0.20 & 0.15 \\
   0.10 & 0.20 \\
   \end{array}
 \right),\\
  &M(1)=\left(
   \begin{array}{cc}
   0.15 & 0.20 \\
   0.30 & 0.40 \\
   \end{array}
 \right),
M(2)=\left(
   \begin{array}{cc}
   0.25 & 0.12 \\
   0.20 & 0.14 \\
   \end{array}
 \right),\\
&M(3)=\left(
   \begin{array}{cc}
   0.15 & -0.20 \\
   -0.23 & 0.22 \\
   \end{array}
 \right),
 M(4)=\left(
   \begin{array}{cc}
   0.15 & 0.15 \\
   0.40 & 0.20 \\
   \end{array}
 \right).
 \end{align*}
The neuron activation functions are selected as
\begin{align*}
&f(x(k))=\left(
   \begin{array}{cc}
   \tanh(0.03x_1(k))\\
   \tanh(0.02x_2(k))\\
   \end{array}
 \right).
\end{align*}
It is readily seen that there exist matrices
\begin{align*}
&F_1=\left(
   \begin{array}{cc}
   0.2   & 0\\
   0  &  0.1\\
   \end{array}
 \right),
 F_2={
 \left(
   \begin{array}{cc}
   0.1  &  0\\
   0  &  0.2\\
   \end{array}
 \right)},
\end{align*}
such that Assumption 1 holds.

Let the transition probability matrix be
\begin{align*}
\pi=
 \left(
   \begin{array}{cccc}
   0.3  &  0.2 & 0.1 & 0.4\\
   0.3  &  0.2 & 0.3 & 0.2\\
   0.1  &  0.1 & 0.5 & 0.3\\
   0.2  &  0.2 & 0.1 & 0.5\\
   \end{array}
 \right),
\end{align*}
 and the exogenous disturbance $w(k)=e^{-0.05^k}\sin(k),v(k)=2e^{-0.05^k}\cos(2k)$.

In the simulations, we obtain a series of feasible solutions of the gain matrices as follows
\begin{align*}
&K_{11}=10^{-4}\times{
 \left(
   \begin{array}{cc}
    0 &  0.3587\\
  0 & 0.1064\\
   0  & 0.0002\\
   0.0001 &  0\\
   \end{array}
 \right)},
 K_{12}=10^{-4}\times{
 \left(
   \begin{array}{cc}
     -0.0036 & 0\\
  0.1864  &  0.1317\\
   0 &  0\\
  0.1321 & 0\\
   \end{array}
 \right)},\\
 &K_{21}=10^{-5}\times{
 \left(
   \begin{array}{cc}
                   0  &  0.9012\\
         0 &  -0.5550\\
        0 &   -0.0034 \\
         -0.0046 &   0\\
   \end{array}
 \right)},
 K_{22}=10^{-3}\times{
 \left(
   \begin{array}{cc}
       0.2765 &  0\\
   0.1415  & -0.1157\\
    0 &  0\\
    -0.0002  &  0\\
   \end{array}
 \right)},\\
 &K_{31}=10^{-5}\times{
 \left(
   \begin{array}{cc}
    0 &  -0.6184\\
  0 & 0.2952\\
   0  & 0.0067\\
   -0.0006 &  0\\
   \end{array}
 \right)},
 K_{32}=10^{-4}\times{
 \left(
   \begin{array}{cc}
     0.3930 & 0\\
  0.0150 &  -0.0003\\
   0 &  0\\
  -0.0005 &0\\
   \end{array}
 \right)},\\
 &K_{41}=10^{-4}\times{
 \left(
   \begin{array}{cc}
                   0  &  -0.6830\\
         0 &  0.6758\\
        0 &   0.0046 \\
         0.0049 &   0\\
   \end{array}
 \right)},
 K_{42}=10^{-4}\times{
 \left(
   \begin{array}{cc}
       0.0735 &  0\\
   -0.1131 &  -0.0398\\
    0 &  0\\
    -0.0402  &  0\\
   \end{array}
 \right)}.
\end{align*}

\begin{figure}[h]
  \centering
  \includegraphics[width=0.75\textwidth]{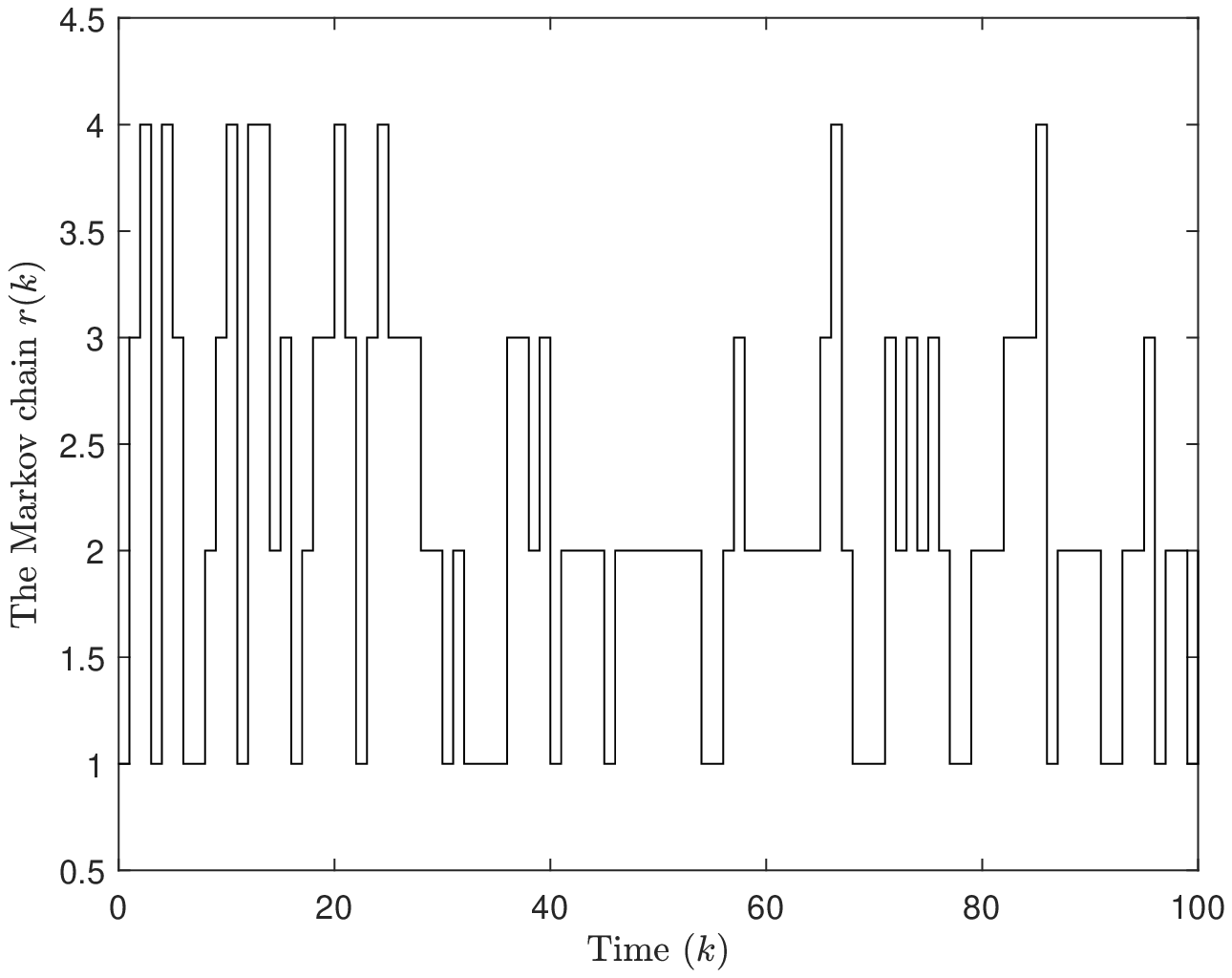}\\
  \caption{Markov switching between four modes}\label{fig:1}
\end{figure}

\begin{figure}[h]
  \centering
  \includegraphics[width=0.75\textwidth]{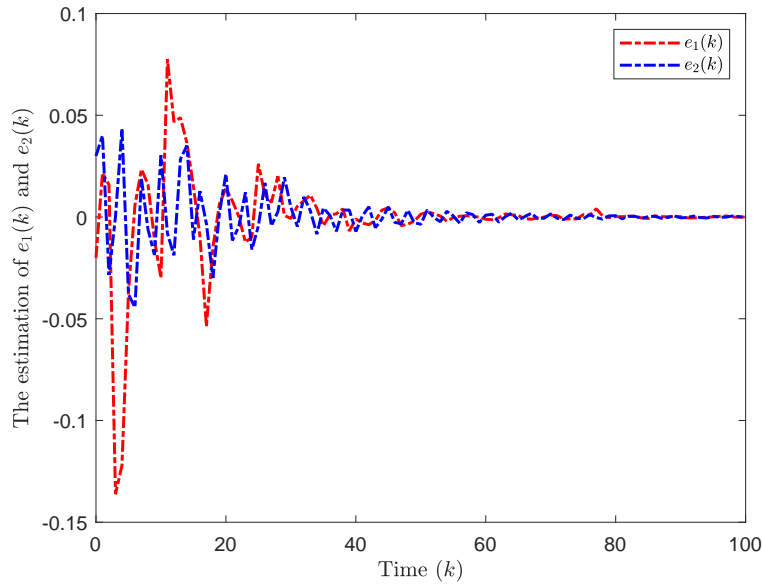}\\
  \caption{Estimation error $e_1(k)$ and $e_2(k)$.}\label{fig:2}
\end{figure}

\begin{figure}[h]
  \centering
  \includegraphics[width=0.75\textwidth]{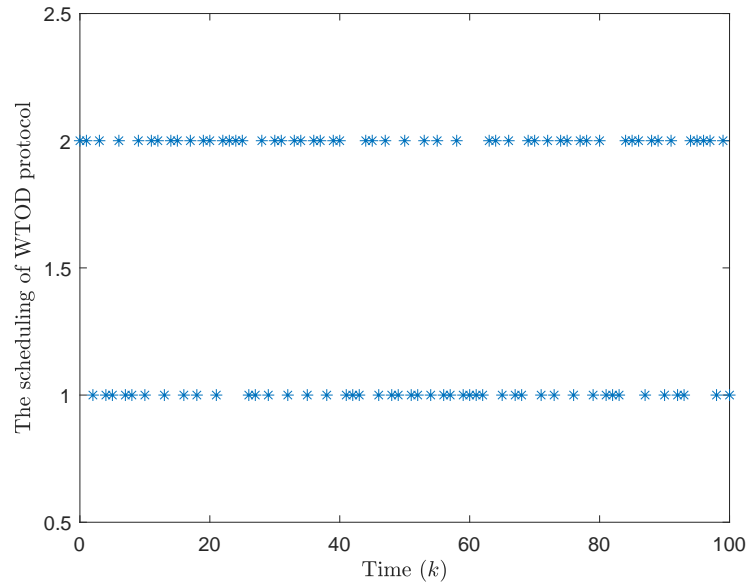}\\
  \caption{The scheduling of WTOD protocol.}\label{fig:3}
\end{figure}

Figs. \ref{fig:1}-\ref{fig:3} show the simulation results for the model. Figure \ref{fig:1} shows the evolution process of the Markov chain.  The results for the estimation error are shown in Figs. \ref{fig:2}. Fig. \ref{fig:3} depicts the number of sensor nodes MJNNs under WTOD protocol. It is readily observed from the simulation results that the designed estimator is effective.
%\begin{remark}
%In a given example, which has two modes and two sensor nodes, Table. \ref{tab:1} represents the update number of two-state output nodes with and without the scheduling of the Round-Robin protocol, respectively. The red and blue circles in Fig. \ref{fig:7} respectively denote that the number of two output signals $y_1(k)$, $y_2(k)$ switch to select two sensor nodes under the protocol schedule. From the comparison, we can know that under the Round-Robin protocol, the transmission burden is greatly reduced and resources are saved.
%\end{remark}

%\begin{figure}[h]
%  \centering
%  % Requires \usepackage{graphicx}
%  \includegraphics[width=0.75\textwidth]{fig3}\\
%  \caption{State $x_1(k)$ and its estimate $\hat{x}_1(k)$.}\label{fig:3}
%\end{figure}
%\begin{figure}[h]
%  \centering
%  % Requires \usepackage{graphicx}
%  \includegraphics[width=0.75\textwidth]{fig4}\\
%  \caption{State $x_2(k)$ and its estimate $\hat{x}_2(k)$.}\label{fig:4}
%\end{figure}
%
%
%\begin{table}\caption{Update number of measurement output.}\centering\label{tab:1}
%\begin{tabular}{c|cc}
%  \hline
%  % after \\: \hline or \cline{col1-col2} \cline{col3-col4} ...
%  Measurement output nodes & $y_1$ & $y_2$ \\
%   \hline
%  Transmission number without RRP & 101 & 101 \\
%  Transmission number with RRP & 50 & 50 \\
%  \hline
%\end{tabular}
%\end{table}
%\begin{figure}[h]
%  \centering
%  % Requires \usepackage{graphicx}
%  \includegraphics[width=0.75\textwidth]{fig7}\\
%  \caption{Update of measurement output.}\label{fig:7}
%\end{figure}

\section{Conclusions}\label{sec:5}

\end{document}